\documentclass[aps,prd,a4paper,nosuperscriptaddress,
twocolumn,
nofootinbib,showpacs,showkeys,amsfonts,amssymb,amsmath]{revtex4-1}

\usepackage{amssymb,latexsym}
\usepackage{amsmath, amsthm}
\usepackage{amscd}
\usepackage{times}
\usepackage{epsfig}
\usepackage{psfrag}
\usepackage{graphicx}
\newcommand{\bea}{\begin{eqnarray}}
\newcommand{\eea}{\end{eqnarray}}
\newcommand{\be}{\begin{equation}}
\newcommand{\ee}{\end{equation}}

\begin{document}

\title[]{Note on the effect of a massive accretion disk in the measurements of black hole spins}

\author{Cosimo Bambi} 
\email{bambi@fudan.edu.cn} 
\affiliation{Center for Field Theory and Particle Physics and Department of Physics,
Fudan University, 220 Handan Road, 200433 Shanghai, China}

\author{Daniele Malafarina} 
\email{daniele@fudan.edu.cn} 
\affiliation{Center for Field Theory and Particle Physics and Department of Physics,
Fudan University, 220 Handan Road, 200433 Shanghai, China}

\author{Naoki Tsukamoto} 
\email{tsukamoto@fudan.edu.cn} 
\affiliation{Center for Field Theory and Particle Physics and Department of Physics,
Fudan University, 220 Handan Road, 200433 Shanghai, China}

\begin{abstract}
The spin measurement of black holes has important 
implications in physics and astrophysics. Regardless of the specific technique 
to estimate the black hole spin, all the current approaches assume that the 
space-time geometry around the compact object is exactly described by the 
Kerr solution. This is clearly an approximation, because the Kerr metric is a 
stationary solution of the vacuum Einstein equations. In this paper, we estimate 
the effect of a massive accretion disk in the measurement of the black hole spin 
with a simple analytical model. For typical accretion disks, the mass of the disk 
is completely negligible, even for future more accurate measurements. However, 
for systems with very massive disks the effect may not be ignored.
\end{abstract}

\pacs{97.60.Lf, 04.40.-b, 04.20.Jb}

\maketitle

\section{Introduction}

In 4-dimensional general relativity, an uncharged black hole (BH) is described 
by the Kerr solution and it is characterized by two parameters, associated respectively 
with the mass $M$ and the spin angular momentum $J$ of the object~\cite{kerr}. 
Astrophysical BHs are thought to be well described by the Kerr solution. Initial 
deviations from the Kerr background can be quickly radiated away through the 
emission of gravitational waves~\cite{price}. An initially non-vanishing electric 
charge can be soon neutralized in the highly ionized environment of the 
BH~\cite{bdp}. The presence of the accretion disk is typically completely negligible.

The BH mass can be measured with dynamical methods, by studying the orbital 
motion of individual stars. This kind of measurement is robust and one can use 
Newtonian mechanics, because the stars are far from the BH. The estimate of the 
spin is much more challenging. The spin has no gravitational effects in the 
Newtonian theory, and therefore it can be measured only probing the space-time 
geometry close to the BH. At present, there are two relatively reliable techniques 
to measure BH spins: the continuum-fitting method~\cite{cfm} and the analysis 
of the K$\alpha$ iron line~\cite{iron}. The continuum-fitting method is based on 
the study of the thermal spectrum of geometrically thin and optically thick accretion 
disks. It can be used only for stellar-mass BHs because the disk temperature is 
proportional to $M^{-1/4}$ and the spectrum turns out to be in the X-ray range 
for stellar-mass BHs and in the UV/optical range for super-massive BHs. In the 
latter case, absorption from dust makes an accurate measurement impossible. The 
analysis of the K$\alpha$ iron line can be applied to both stellar-mass and 
super-massive BHs and is based on the study of the profile of this line: the latter 
is intrinsically narrow in frequency, while the one in the spectrum of BHs is affected 
by special and general relativistic effects.

Both the continuum-fitting method and the analysis of the iron line strongly rely 
on the assumption that the inner edge of the accretion disk is at the innermost 
stable circular orbit (ISCO) of the background metric. In an exact Kerr metric, the 
ISCO radius in Boyer-Lindquist coordinates $\{t,r,\theta,\phi \}$ is given by~\cite{bpt}
\bea
r_{\rm ISCO} &=& \left[ 3 + Z_2 \mp \sqrt{(3 - Z_1)(3 + Z_1 + 2 Z_2)} \right] M \, , \\
Z_1 &=& 1 + \left(1 - a_*^2\right)^{1/3}
\left[\left(1 + a_*\right)^{1/3} + \left(1 - a_*\right)^{1/3} \right] \, , \nonumber\\
Z_2 &=& \sqrt{3 a_*^2 + Z_1^2} \, , \nonumber
\eea
where $a_* = a/M = J/M^2$ is the spin parameter and the sign $-$ $(+)$ is for
corotating (counterrotating) orbits. For a Schwarzschild BH ($a_* = 0$), the ISCO 
radius is at $r_{\rm ISCO} = 6\,M$ and its value decreases (increases) for a rotating 
BH and a corotating (counterrotating) disk.

In general relativity, the radial coordinate has not really any physical meaning, 
being determined by the coordinate system, which is arbitrary. In the analysis of 
the thermal spectrum of thin disks, one can see that the key-quantity is the radiative 
efficiency in the Novikov-Thorne model~\cite{bb}. In the case of the iron line profile,
the picture is more complicated, but the value of the radiative efficiency in the 
Novikov-Thorne model can still be used as a crude estimate to compare different
spacetimes with similar iron lines~\cite{cc}. In the Kerr metric, the specific 
energy of a test-particle on the equatorial circular orbit of radius $r$ is
\bea\label{eq-ee}
E = \frac{r^{3/2} - 2 M r^{1/2} \pm a M^{1/2}}{r^{3/4} 
\sqrt{r^{3/2} - 3 M r^{1/2} \pm 2 a M^{1/2}}} \, .
\eea
The Novikov-Thorne radiative efficiency is
\bea
\eta_{\rm NT} = 1 - E_{\rm ISCO} \, ,
\eea
where $E_{\rm ISCO}$ is the energy in Eq.~(\ref{eq-ee}) evaluated at $r=r_{\rm ISCO}$.
In the slow-rotating case $(a\ll M)$, the specific conserved energy of a test-particle
at the ISCO radius is
\begin{eqnarray}\label{eq:E_ISCO_slow}
E_{\rm ISCO}=\frac{2\sqrt{2}}{3} \mp \frac{\sqrt{3}}{54} \, \frac{a}{M}.
\end{eqnarray}

Here we attempt to estimate how the mass of the accretion disk can affect the 
above measurement. We make use of an analytical model for the BH 
plus disk system to show that there exists a degeneracy in the measurement 
of the BH spin angular momentum when the disk is sufficiently massive with respect 
to the mass of the central object. The effect of massive disks on gravitational 
wave measurements was previously studied in~\cite{enricogw}.

\section{Axially symmetric spacetimes}\label{weyl}

In the following we review the formalism to  obtain an exact solution describing a system composed by a non-rotating BH plus a thin disk of matter.
In Weyl coordinates $\{t,\rho,z,\phi\}$, the general form of a static and axially 
symmetric metric depends only upon two functions $\lambda(\rho,z)$ and $\nu(\rho,z)$ and is given by
\be\label{Weyl}
ds^2=-e^{2\lambda}dt^2+e^{2(\nu-\lambda)}(d\rho^2+dz^2)
+\rho^2e^{-2\lambda}d\phi^2 \, .
\ee
With the line element in Eq.~(\ref{Weyl}), the Einstein equations become
\bea
4\pi e^{2\nu-2\lambda}(T^\phi_\phi-T^t_t)&=&\nabla^2\lambda=
\lambda_{,\rho\rho}+\frac{\lambda_{,\rho}}{\rho}+\lambda_{,zz} \, , \\
4\pi (T_{\rho\rho}-T_{zz})&=& \frac{\nu_{,\rho}}{\rho}
- (\lambda_{,\rho})^2+(\lambda_{,z})^2 \, , \\
4\pi T_{\rho z}&=& \frac{\nu_{,z}}{2\rho} -\lambda_{,\rho}\lambda_{,z} \, \\
4\pi e^{2\nu-2\lambda}(T^\phi_\phi+T^t_t)&=& (\lambda_{,\rho})^2
+(\lambda_{,z})^2+\nu_{,\rho\rho}+\nu_{,zz}-\nabla^2\lambda \, , \nonumber\\
\eea
and in vacuum they reduce to
\bea \label{laplace}
&&\nabla^2\lambda=\lambda_{,\rho\rho}+
\frac{\lambda_{,\rho}}{\rho}+\lambda_{,zz}=0 \, , \\ 
&&\nu_{,z}=2\rho\lambda_{,\rho}\lambda_{,z} \, \qquad 
\nu_{,\rho}=\rho\left[ (\lambda_{,\rho})^2-(\lambda_{,z})^2 \right] \, .
\label{eq:nu_rho}
\eea

Notice that Eq.~\eqref{laplace} is just the Laplace equation in the flat two-dimensional space. Therefore, in principle, every vacuum solution is known. In fact, once we choose a solution $\lambda$ of Eq.~\eqref{laplace} for the exterior region of a certain Newtonian density distribution, we use Eqs.~\eqref{eq:nu_rho} to find $\nu$ and obtain the corresponding solution for the vacuum Einstein equations. 
Moreover, since Eq.~\eqref{laplace} is linear, if $\lambda_1$ and $\lambda_2$ are two solutions then also $\lambda=\lambda_1+\lambda_2$ must be a solution. 
The non linearity of Einstein equations comes from the second function $\nu$ which is then given by
\begin{eqnarray}
 \label{nu}
\nu=\nu_1+\nu_2+2\int\rho\left[(\lambda_{1,\rho}\lambda_{2,\rho}
-\lambda_{1,z}\lambda_{2,z})d\rho \right. \nonumber\\
\left. +(\lambda_{1,\rho}\lambda_{2,z}+\lambda_{1,z}\lambda_{2,\rho})dz\right] \, ,
\end{eqnarray}
where $\nu_1$ and $\nu_2$ are obtained from Eq.~(\ref{eq:nu_rho}), 
respectively for $\lambda_1$ and $\lambda_2$.

Spherical symmetry is a subcase of axial symmetry and therefore it must be possible to obtain the Schwarzschild BH in the above formalism. In Weyl coordinates, the Schwarzschild solution is given by a function $\lambda_{\rm BH}$ corresponding to a Newtonian source 
of constant density $\omega=1$ distributed along the $z$ axis, from $z=-M$ to 
$z=M$. Then $\lambda_{\rm BH}$ and $\nu_{\rm BH}$ are given by
\bea \label{eq:lambda_Sch}
\hspace{-0.4cm}
\lambda_{\rm BH}&=&\frac{1}{2}
\ln \left(\frac{K_+ + K_- - 2M}{K_+ + K_- + 2M}\right) \, , \\
\hspace{-0.4cm}
\nu_{\rm BH} &=& \frac{1}{2}\ln \frac{\left(K_{-}+K_{+}-2M\right)
\left(K_{-}+K_{+}+2M\right)}{4K_{-}K_{+}} \, ,
\eea
where $K_{\pm}=\sqrt{\rho^2+(z\pm M)^2}$. For $z=0$, $\nu_{\rm BH}$ 
reduced to
\begin{eqnarray}
\nu_{\rm BH}
=\frac{1}{2}\ln \frac{\rho^{2}}{\rho^{2}+M^{2}} \, .
\end{eqnarray}
The Schwarzschild solution in the usual Schwarzschild coordinates 
$\{t,r,\theta,\phi \}$ is recovered after the following change of coordinates
\be
\rho=\sqrt{r^2-2Mr}\sin\theta\; , \; \; z=(r-M)\cos\theta \, .
\ee

A realistic thin disk composed of ordinary matter was obtained by Lemos and Letelier in~\cite{Lemos:1993qp} by making an inversion of a thin disk of the Morgan-Morgan family \cite{Morgan}. The disk has an inner edge at $b$ and the metric function $\lambda_{\rm D}$ solution of Eq.~\eqref{laplace} is given by a Newtonian 
density distribution $\omega$ concentrated on the equatorial plane as
\be
\omega(\rho, z)=\frac{2mb}{\pi^2\rho^4}\sqrt{\rho^2-b^2} \delta(z) \, .
\ee 
The disk is located on the equatorial plane $z=0$ and it can be described in a distributional sense with the thin shell formalism that was developed by Israel in \cite{Israel}. Then the full four dimensional metric is divided into three parts: $(i)$ a vacuum 4-dimensional part above the disk, $ds^2_+$, 
for $z>0$, $(ii)$ a vacuum 4-dimensional part below the disk, $ds^2_-$, for $z<0$ 
(the same as the above with negative $z$), $(iii)$ a 3-dimensional part with 
non-vanishing energy-momentum tensor, $ds^2_\Sigma$, restricted to the hypersurface $\Sigma$ given by $z=0$. The continuity of 
the first fundamental form ensures that the metric is continuous across $z=0$. 
The jump of the second fundamental form across $\Sigma$ is related to the matter content 
of the disk. The energy-momentum tensor is restricted to the 
equatorial plane and can be understood as a delta-like 
distribution on $z=0$. It is given by
\bea \label{eq:T_tt}
\epsilon&=&-T^t_t=4e^{2\lambda_{\rm D}-2\nu_{\rm D}}(1-\rho\lambda_{{\rm D},\rho})\lambda_{{\rm D},z}
\delta(z) \; , \\ \label{eq:T_phiphi}
p_{\phi\phi}&=&T^\phi_\phi= 4e^{2\lambda_{\rm D}-2\nu_{\rm D}} \rho \lambda_{{\rm D},\rho} 
\lambda_{{\rm D},z}\delta(z)  \; , \\ \label{eq:T_rhorho}
T^\rho_\rho&=&T^z_z=0\; .
\eea

We can introduce oblate spheroidal coordinates $\{t,R,\theta,\phi\}$ that are related to the Weyl coordinates by
\be \label{eq:oblate}
\rho=\sqrt{R^2+b^2}\sin\theta\; , \; \; z=R\cos\theta \; .
\ee
Then $\lambda_{\rm D}$ is given 
by (see \cite{Semerak_Zellerin_Zacek_1999} for details)
\begin{widetext}
\bea \label{eq:lambda_disk}
\lambda_{\rm D}&=&-\frac{m}{\pi (R^2+b^2\sin^2\theta)^{3/2}}\left\{\frac{(2R^2-b^2\sin^2\theta)b|\cos\theta|}{\sqrt{R^2+b^2\sin^2\theta}}\left[\frac{b|\cos\theta|}{\sqrt{R^2+b^2\sin^2\theta}}\cot^{-1}\left(\frac{b|\cos\theta|}{\sqrt{R^2+b^2\sin^2\theta}}\right)-1\right]\right.\\ \nonumber
&&\left.+(2R^2+b^2\sin^2\theta)\cot^{-1}\left(\frac{b|\cos\theta|}{\sqrt{R^2+b^2\sin^2\theta}}\right) \right\}.
\eea
\end{widetext}

\begin{figure}
\begin{center}
\includegraphics[type=pdf,ext=.pdf,read=.pdf,width=8.0cm]{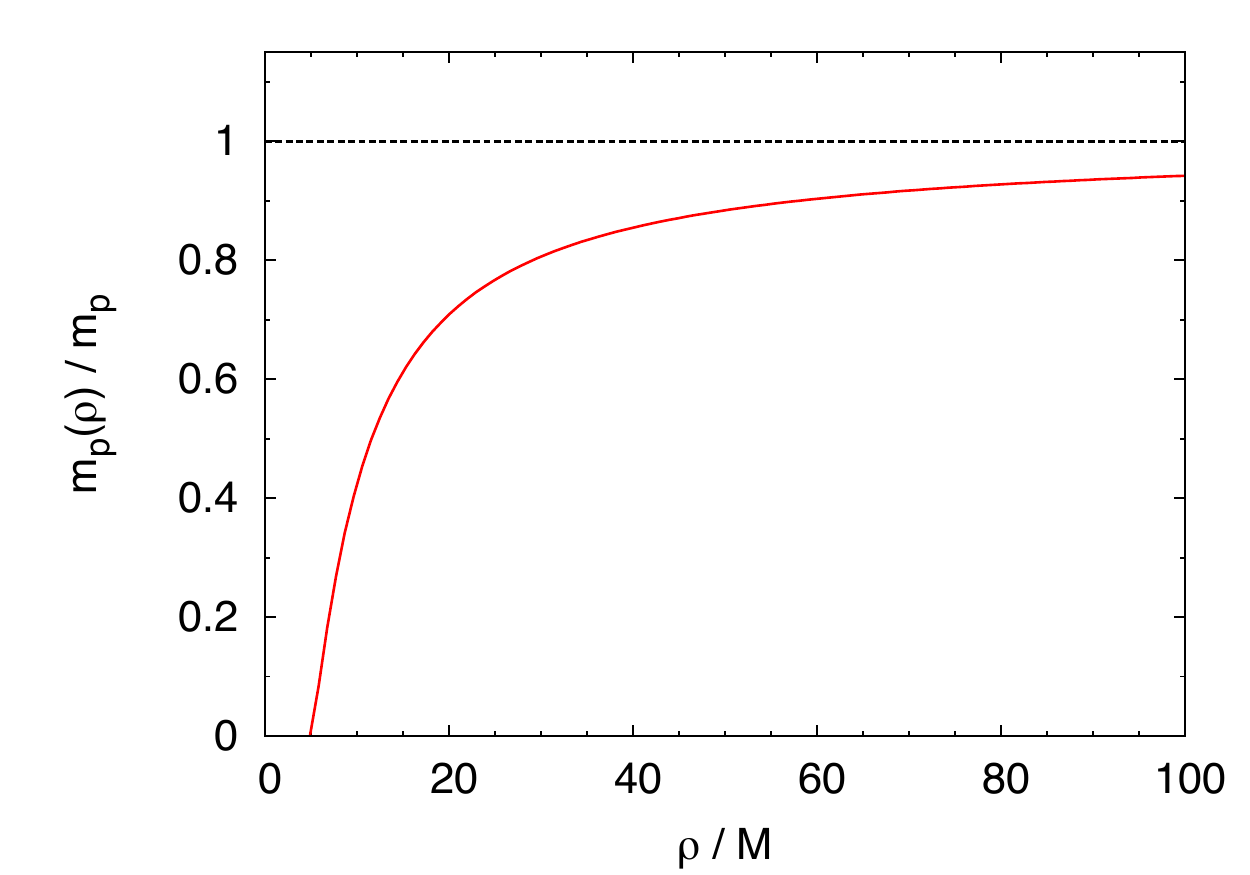}
\end{center}
\caption{Fraction of the proper mass of the disk within the radius $\rho$.}
\label{fig1}
\end{figure}

A system composed by a BH surrounded by a thin accretion disk can be described by an exact solution of the Weyl class where the
line element in Eq.~(\ref{Weyl}) has
$\lambda=\lambda_{\rm BH}+\lambda_{\rm D}$,
while $\nu$ is solution of Eq.~\eqref{nu}. The metric of the space-time depends upon
three parameters: the BH mass $M$, the disk mass $m$, and the disk's inner radius 
$b$. The proper mass of the disk within the radial coordinate $\rho$ is given by
\begin{widetext}
\begin{eqnarray}
\hspace{-0.3cm}
m_{p}(\rho)=
32mb\int^{\rho}_{b} \frac{\sqrt{\rho'^{2}-b^{2}} 
\left( \sqrt{\rho'^{2}+M^{2}}+M \right)^{2}}{\rho'^{5}}
\left[ 1-\frac{M}{\sqrt{\rho'^{2}+M^{2}}}-
\frac{(2\rho'^{2}-3b^{2})m}{2\rho'^{3}} \right] 
\exp \left[ \frac{(2\rho'^{2}-b^{2})m}{2\rho'^{3}} \right] d\rho' \, .
\end{eqnarray}
\end{widetext}
The proper mass $m_{p}(\rho)$ does depend on the BH mass $M$ because it 
includes the gravitational binding energy. Fig.~\ref{fig1} shows the fraction of the
disk mass, $m_{p}(\rho)/m_p$ where $m_p = m_p(\infty)$, as a function of the 
radial coordinate $\rho$. In the Lemos-Leteilier disk, most of the mass is concentrated 
at small radii, not far from the BH.

\section{ISCO radius in presence of a massive disk and implications on the 
measurement of the black hole spin}

If we plug the functions $\lambda_{\rm BH}$, $\nu_{\rm BH}$, $\lambda_{\rm D}$, 
and $\nu_{\rm D}$ into Eq.~\eqref{nu}, we get the $\nu$ function for the spacetime
of the BH surrounded by the accretion disk.
On the equatorial plane $z=0$, one finds
\begin{widetext}
\begin{eqnarray}
\nu
&=&\frac{1}{2}\ln \frac{\rho^{2}}{\rho^{2}+M^{2}}
+\frac{m}{M}\left[ 2-\frac{2\sqrt{\rho^{2}+M^{2}}}{\rho}+\frac{2b^{2}}{M^{2}}
-\frac{3b^{2}\sqrt{\rho^{2}+M^{2}}}{M^{2}\rho}+\frac{b^{2}(\rho^{2}+M^{2})^{3/2}}{M^{2}\rho^{3}} \right] \nonumber\\
&&+m^{2} \left[ -\frac{1}{2\rho^{2}} +\left( \frac{3}{4}+\frac{4}{\pi^{2}} \right) \frac{b^{2}}{\rho^{4}} 
-\left( \frac{3}{8}+\frac{8}{3\pi^{2}} \right) \frac{b^{4}}{\rho^{6}} \right].
\end{eqnarray}
\end{widetext}

From the conservation of the rest-mass, the
motion of a test particle on the equatorial plane is governed by the equation
\begin{eqnarray}
\dot{\rho}^{2} + V_{\rm eff}=0 \, ,
\end{eqnarray}
where 
\begin{eqnarray}
V_{\rm eff}= \frac{1}{g_{\rho\rho}}\left( 1+\frac{E^{2}}{g_{tt}}
+\frac{L^{2}}{g_{\phi\phi}} \right) \; ,
\end{eqnarray}
is the effective potential, $E =-g_{tt}\dot{t}$ is the specific conserved energy, and
$L=g_{\phi\phi}\dot{\phi}$ is the specific angular momentum of the particle. Here, 
a dot $\dot{}$ indicates the derivative with respect to the proper time for the particle.
By definition, circular orbits are located at the zeros and the turning points of the effective 
potential: $\dot{\rho} = 0$, which implies $V_{\rm eff}=0$, and $\ddot{\rho}=0$,
requiring $V_{{\rm eff},\rho}=0$.
From these conditions, one finds
\begin{eqnarray}\label{eq:E_c}
\hspace{-0.5cm}
E= \sqrt{\frac{e^{2\lambda}(1-\lambda_{,\rho}\rho)}{1-2\lambda_{,\rho}\rho}} \, , \quad
L = \sqrt{\frac{\lambda_{,\rho}\rho^{3}}{e^{2\lambda}(1-2\lambda_{,\rho}\rho)}} \, .
\end{eqnarray}

The ISCO radius is at the minimum of $E$ and $L$ and can be inferred from
\begin{eqnarray}\label{eq:E_crho=0}
\lambda_{,\rho\rho}\rho+\lambda_{,\rho}\left[ 4(\lambda_{,\rho})^{2}\rho^{2}-6\lambda_{,\rho}\rho+3 \right]=0.
\end{eqnarray}
For $m\ll M$ and $mb^{2}\ll M^{3}$, the radius of the ISCO is
\begin{eqnarray}\label{eq:isco}
\rho_{\rm ISCO} \approx 2\sqrt{6}M - \frac{25 M}{1536} 
\frac{73b^{2}-368M^{2}}{M^2} \, \frac{m}{M} \, .
\end{eqnarray}
The corresponding specific energy for a test-particle at the ISCO radius is
\begin{eqnarray}\label{eq:E_ISCO_BH_DISK}
E_{\rm ISCO} \approx \frac{2\sqrt{2}}{3} 
\left( 1-\frac{\sqrt{6}}{4608}\frac{17b^{2}-16M^{2}}{M^2} \, \frac{m}{M}\right) \, .
\end{eqnarray}

\begin{figure*}
\begin{center}
\includegraphics[type=pdf,ext=.pdf,read=.pdf,width=7.2cm]{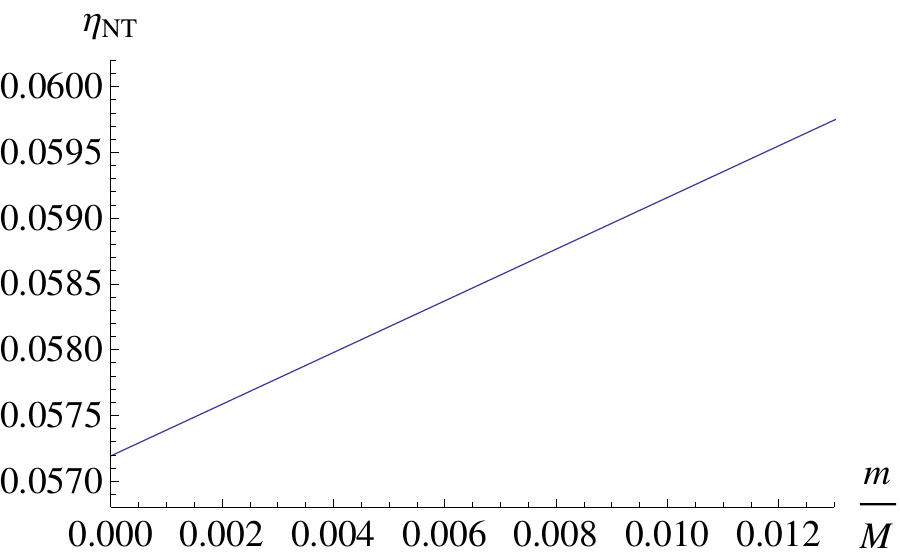}
\hspace{1.4cm}
\includegraphics[type=pdf,ext=.pdf,read=.pdf,width=7.2cm]{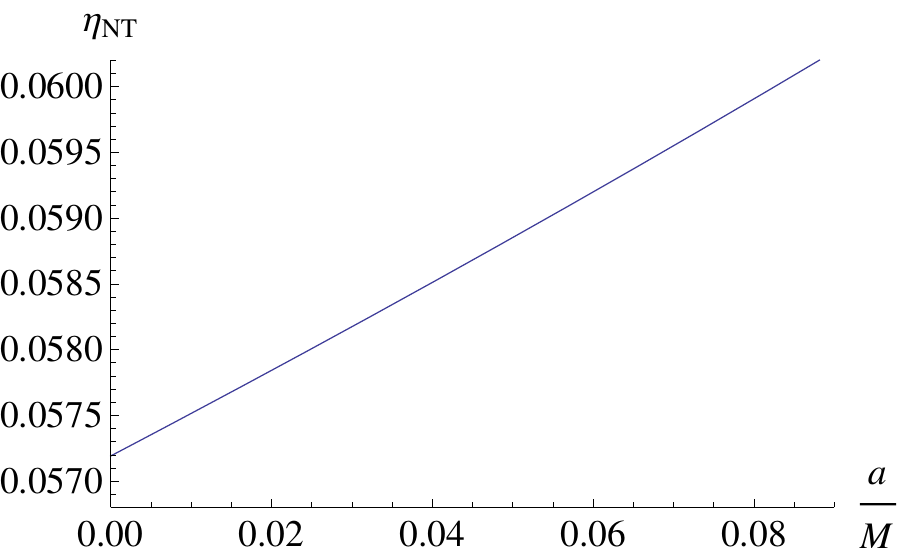}
\end{center}
\caption{Novikov-Thorne radiative efficiency $\eta_{\rm NT} = 1 - E_{\rm ISCO}$ as
a function of the disk to BH mass ratio $m/M$ in the case of a massive disk around
a non-rotating BH (left panel) and as a function of the spin parameter $a/M$ in the
case of zero-mass disk around a rotating BH (right panel).}
\label{fig2}
\end{figure*}

Assuming that the inner edge of the disk $b$ is at the ISCO radius $\rho_{\rm ISCO}$,
one finds
\begin{eqnarray}
\rho_{\rm ISCO} \approx 2\sqrt{6}M-\frac{4325 M}{192} \, \frac{m}{M} \, ,
\end{eqnarray} 
and the Novikov-Thorne radiative efficiency in presence of a massive disk is
\bea\label{eq-eta-ntm}
\eta_{\rm NT} \approx 1 - \frac{2\sqrt{2}}{3} 
\left( 1-\frac{49\sqrt{6}}{576} \, \frac{m}{M} \right) \, .
\eea
Eq.~\eqref{eq-eta-ntm} has to be compared with Eq.~\eqref{eq:E_ISCO_slow}: 
the effects of a massive disk can mimic the ones of a rotating BH with a 
massless corotating disk. In particular,
\begin{eqnarray}
\frac{a}{M} \approx \frac{49}{8} \, \frac{m}{M} \, .
\end{eqnarray}
Fig.~\ref{fig2} shows $\eta_{\rm NT}$ as a function of $m/M$ (left panel) and 
$a/M$ (right panel). Typical accretion disks around stellar-mass BHs have 
$m/M \sim 10^{-9} - 10^{-10}$ and their mass is not concentrated as close as to the
compact object as shown in Fig.~\ref{fig1}. Considering that, even in the most 
favorable conditions, a spin measurement can be at the level of $a/M = 0 \pm 0.05$, 
the effect of the mass of the accretion disk can be definitively neglected, even in 
the case of future more accurate measurements. However, BHs with very massive 
disks may exist, and when $m/M$ approaches 0.01, the effect may not be ignored.

\begin{acknowledgments}
This work was supported by the NSFC grant No.~11305038, 
the Shanghai Municipal Education Commission grant for Innovative 
Programs No.~14ZZ001, the Thousand Young Talents Program, 
and Fudan University.
\end{acknowledgments}

\end{document}